\documentclass[aps,pre,showpacs,showkeys,twocolumn,superscriptaddress,
floats,floatfix,preprintnumbers,longbibliography]{revtex4-1}
\usepackage{amssymb,amsmath}
\usepackage{bm,url,graphics,subfigure,epsfig,rotating,float}
\usepackage{mathrsfs} 
\usepackage{hyperref}
\usepackage{color,soul,graphicx,dcolumn,enumerate}
\usepackage{ulem}
\graphicspath{{./}{./Figs/}}
\def \Lint {L_{\rm I}}
\def \taua {\tau_{\ast}}
\def \uek {\bm{u}_{\rm 1}}
\def \udo {\bm{u}_{\rm 2}}
\def \utin {\bm{u}_{\rm 3}}
\def \uchar {\bm{u}_{\rm 4}}

\def \Sp  {S_{\rm p}}

\def \mP {\mathcal{P}}

\def \kmax {k_{\rm max}}

\def  \xx  {{\bm x}}
\def \rr   {\bm{r}}

\def \dive {{\bm \nabla}\cdot}

\def \delt {\partial_t}
\def \delR {\partial_R}
\def \Rnot {R_{\rm 0}}

\newcommand{\avg}[1]{\left\langle #1\right\rangle}

\def \LI {{\mathcal L}_{\rm I}}
\def \TL {T_{\rm L}}

\def \kc  {k_{\rm c}}

\def \urms  {u_{\rm rms}}

\def \Np  {N_{\rm p}}

\def \zetap {\zeta_{\rm p}}

\newcommand{\eq}[1]{~(\ref{#1})}

\newcommand{\Fig}[1]{Fig.~(\ref{#1})}
\newcommand{\subfig}[2]{Fig.~(\ref{#1}#2)}

\newcommand{\bfig}{\begin{figure}}
\newcommand{\efig}{\end{figure}}
\newcommand{\bc}{\begin{center}}
\newcommand{\ec}{\end{center}}
\newcommand{\bea}{\begin{eqnarray}}
\newcommand{\eea}{\end{eqnarray}}

\def \pre {Phys. Rev. E}
\def \prl {Phys. Rev. Lett.}
\def \uvec {\bm{u}}
\def \xvec {\bm{x}}
\def \fvec {\bm{f}}
\def \kvec {\bm{k}}
\def \Xvec {\bm{X}}
\def \urms {u_{\rm rms}}

\def \zetap {\zeta_p}

\def \tauD {\tau_{\rm D}}
\def \tauH {\tau_{\rm H}}
\def \omD {\omega_{\rm D}}
\def \omH {\omega_{\rm H}}

\def \chiDp {\chi^{\rm D}_{\rm p}}
\def \chiHp {\chi^{\rm H}_{\rm p}}

\def \kapDp {\kappa^{\rm D}_{\rm p}}

\def \kapHp {\kappa^{\rm H}_{\rm p}}
\def \chiHp {\chi^{\rm H}_{\rm p}}
\def \chiDp {\chi^{\rm D}_{\rm p}}

\def \PLams {\mathcal{P}_{\Lambda,s}}
\def \PLamH {\mathcal{P}_{\Lambda,{\rm H}}}

\def \dRV  {\delta_{\rm R}V}
\def \eps  {\varepsilon}
\def \tcor {t_{\rm cor}}
\def \Pone {\mathcal{P}_{\rm 1}}
\def \Ptwo {\mathcal{P}_{\rm 2}}
\def \Aone {A_{\rm 1}}
\def \Atwo {A_{\rm 2}}
\def \wone {w_{\rm 1}}
\def \wtwo {w_{\rm 2}}
\def \wthA {w_{\rm 3A}}
\def \wthB {w_{\rm 3B}}
\def \fhat {\hat{\bm{f}}}

\def \dup   {\delta u_{\parallel}}
\newcommand{\SMat}[1]{Appendix~\ref{#1}}

\begin{document} 
\title{ Uncovering the multifractality of Lagrangian pair dispersion
in shock-dominated turbulence}
\author{Sadhitro De}
\email{sadhitrode@iisc.ac.in}
\affiliation{Department of Physics, Indian Institute of Science, Bangalore
  560012, India}
\author{Dhrubaditya Mitra}
%DM orcid 0000-0003-4861-8152
\email{dhruba.mitra@gmail.com}
\affiliation{Nordita, KTH Royal Institute of Technology and
Stockholm University, Hannes Alfv\'ens v\"ag 12, 10691 Stockholm, Sweden}
\author{Rahul Pandit}
\email{rahul@iisc.ac.in}
\affiliation{Department of Physics, Indian Institute of Science, Bangalore
  560012, India}
%----------------------
\begin{abstract}
  Lagrangian pair dispersion provides insights into mixing in turbulent flows. 
  By direct numerical simulations (DNS) we show that the statistics of pair dispersion in the randomly forced two-dimensional Burgers equation, which is a typical model of shock-dominated turbulence, is very different from its incompressible counterpart because Lagrangian particles get trapped in shocks. 
  We develop a heuristic theoretical framework that accounts for this — a generalization of the multifractal model — whose prediction of the scaling of Lagrangian exit times agrees well with our DNS.
\end{abstract}
%---------------------
\maketitle
%-------------------
%\paragraph{Introduction:}
The statistics of the relative pair dispersion of Lagrangian
particles (tracers) in turbulent flows has numerous physical
applications~\citep{young1999stirring,fal+gaw+var01,Batchelor1952,
  Sawford_Rev_2001,Salazar_Rev_2009,AtmosDisp2001,OzoneDepl1996,
  OceanDisp2013,benzi2023lectures}. 
\textit{Richardson's law}, a pioneering work in this field, states that, 
in a turbulent fluid, the mean-squared displacement between a pair of tracers,
$\avg{R^2(t)}$, at time $t$, scales as
$\avg{R^2(t)}\sim t^{3}$~\cite{Richardson1926}.
In other words, there is a dynamic exponent $z=3/2$.
Richardson's law can be viewed as a consequence of Kolmogorov's
1941 (K41) scaling theory of turbulence~\citep[see, e.g., Ref.][]{benzi2023lectures}.
It is now well established that the K41 theory is incomplete because it does not
account for intermittency, which arises predominantly because of the most dissipative structures in the flow and leads to multifractality~\citep{Fri96}.
  In brief, intermittency and multifractality in homogeneous and isotropic
  turbulence is characterised by the non-trivial scaling properties of
  moments of velocity differences, $\avg{\delta v (\ell)^p}$, over length scales, $\ell$, i.e., 
  $ \avg{\delta v (\ell)^p} \sim \ell^{\zetap}$,
  with the multiscaling exponent $\zetap$ a nonlinear function of $p$ (K41 yields simple scaling with $\zetap=p/3$)~\footnote{The
  angular brackets denote the average over the turbulent, but
  statistically steady, state of the fluid; and the length scale
  $\ell$ lies between the large length $\Lint$, at which energy is injected,
  and the small length scale $\eta$, at which viscous dissipation becomes
  significant.}.
Therefore, it is natural to hypothesize that Richardson's law must have 
multiscaling corrections too.
In particular, we expect  \textit{dynamic multiscaling}~\citep{lvo+pod+pro97,
  mit+pan03, mit+pan04, ray+mit+pan08, ray+mit+per+pan11},
characterised by not one but  an infinity of dynamic exponents;
specifically, different moments of $R(t)$ should scale with different
powers of $t$. 
However, even the most recent experimental~\citep{ott2000experimental,
  shnapp2018generalization, elsinga2022non, Tan2022PairDisp,
  shnapp2023universal} and numerical~\citep{salazar2009two,bitane2012time,
  bitane2013geometry, bragg2016forward, buaria2015characteristics,
  elsinga2022non} measurements of $\avg{R^2}$ provide only  limited support to
Richardson’s law, because the power-law behavior is observed over a range of scales that covers 
a decade or so. Hence, there seems to be little hope of extracting the
dynamic-multiscaling behavior from experimental or numerical
measurements of moments of $R(t)$. 

It turns out that the \textit{Lagrangian exit time}, conventionally defined as the time taken for the separation between a pair of
tracers~(Lagrangian interval) to exceed a given
threshold (e.g., the \textit{doubling time})~\cite{Bofetta2002_RichardIntermitt,
PairDisp_Bofetta2002,Bofetta1999_Doubling,
FSLE_Artale}, provides a robust measure for dynamic multiscaling. 
From the statistics of these exit times it is possible to extract dynamic multiscaling exponents that match the predictions from the
multifractal model of turbulence ~\cite{benzi1984multifractal,
  parisi1985multifractal,frisch}.

We have, so far, outlined Lagrangian pair dispersion in \textit{incompressible}
turbulence. We turn now to \textit{shock-dominated turbulence}, which refers to highly \textit{compressible} turbulence wherein the total energy in the irrotational modes is significantly larger
than that in the solenoidal modes~\citep{miura1996enstrophy}; this  
poses formidable challenges that we must confront because 
such flows are widely prevalent in many astrophysical systems~\citep{Mordecai2004Review}. 
Over the last two decades, several groups have studied the multifractal properties
of compressible turbulence, ranging from flows that are weakly compressible to those that are shock-dominated~\cite{Schmidt2008Intermit, Konstandin2012EulLag, wang2013cascade,
  wang2013statistics, jagannathan2016reynolds, wang2017spectra,
  donzis2020universality, sakurai2023direct}. 
Additionally, the dynamics of tracers and heavy inertial particles in such systems have also been investigated~\citep{Yang2016LagInt, CompressPl2,Cressman_2004_Surf_EPL,Cressman2003Clustering,Cressman_2004_EulLagSurf}.
However, the possible
multifractal generalization of Richardson's
law and exit time statistics to compressible turbulence has not been established yet.
%---------------------------------
\begin{figure*}
    \centering
    \includegraphics[width=0.95\textwidth]{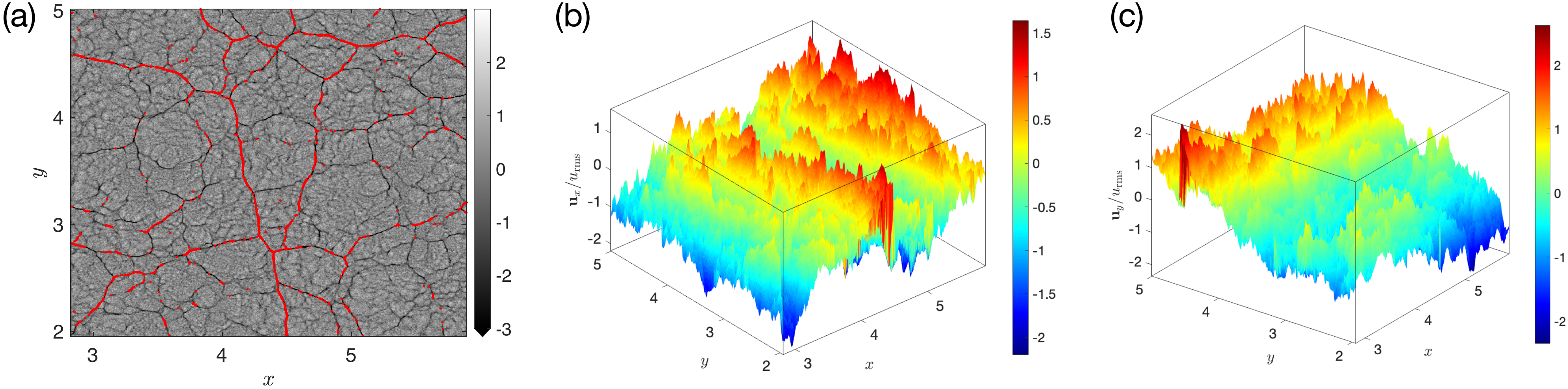}
    \caption{\textbf{Typical snapshots} of a part of our domain in the
      non-equilibrium statistically steady state of turbulence. (a) The divergence of the
      velocity field, $\nabla\cdot\uvec$; shocks are the dark filament-like
      structures with large negative divergences; instantaneous tracer positions
      are marked by red dots. The tracers cluster on the shocks. 
      Surface plots of (b) the $x-$component and (c) the $y-$ component of the velocity
      field $\uvec(\xvec,t)$. The shocks are the large negative 
      jumps. }
    \label{fig:snap}
\end{figure*}
%-------------------------------

We develop the theoretical framework that is required for this
generalization to shock-dominated turbulence. 
It rests on the crucial observation that tracers strongly cluster at the shocks, 
which comprise the most dissipative structures, 
making Lagrangian pair dispersion in such flows qualitatively different from that in incompressible turbulence. 
Consequently, we define two exit times -- \textit{(i)} the \textit{doubling time} and the \textit{(ii) halving time} -- 
as the times taken for a Lagrangian interval to \textit{(a)} go beyond twice and \textit{(b)} shrink below half its initial length, respectively. We carry out explicit direct numerical simulations (DNSs) of
a simplified model of shock-dominated turbulence, namely, the randomly-forced 
two-dimensional (2D) Burgers equation, which is not only rich enough to display the complexities of such turbulence~\cite{Galtier2011Exact,
Schmidt2008Intermit, Federrath2013Universality}, 
but also simple enough for us to develop a heuristic theory for our DNS results. 
We find that the statistics of the doubling times agree with those that
follow from the conventional application of the multifractal model, but those of halving
times do not. 
We generalize this framework to account for the clustering of tracers
on shocks and obtain therefrom the scaling exponents for the moments of the distribution of halving times. 
The results from our heuristic theory are in excellent agreement 
with those from our DNS. 
In a similar vein, we define and investigate the scaling properties of doubling and halving frequencies of Lagrangian intervals, and try to understand them on the basis of the underlying flow intermittency.

The randomly-forced 2D Burgers equation is:
\begin{eqnarray}
  \delt\uvec + \uvec\cdot\nabla\uvec &=&
  \nu\nabla^2\uvec + \fvec(\xvec,t)\,;
  \quad\nabla\times\uvec=0\,; \nonumber \\
        \avg{\fhat(\kvec,t)\cdot\fhat(\kvec',t')}
&\sim& |\kvec|^{-2}\delta(\kvec+\kvec')\delta(t-t')\,.
    \label{eq:Burgers}
\end{eqnarray}
$\uvec(\xvec,t)$ is the Eulerian velocity at position $\xvec$ and time $t$, and
$\nu$ is the coefficient of viscosity. $\fvec(\xvec,t)$ is a
zero-mean, irrotational, Gaussian, white-in-time random 
force whose Fourier components $\fhat(\kvec,t)$ [with $\kvec$ the wave vector] obey the constraint given in \eqref{eq:Burgers}. 
The equal-time and dynamic-scaling properties of the Burgers equation,
which can be mapped to the KPZ equation~\cite{KPZ1}, have been 
studied extensively, but, most often, in one dimension (1D) through DNSs and
renormalization-group  methods~\cite{BURGERS, frisch2001burgulence,
  bec_frisch_khanin_2000, Bec2007BurgersT,BecPRL, PolyakovRG, MedinaRG,
  Chekhlov,HayotStrFn, HayotMultifrac,HayotScaling, BurgMitra,De_DynScal}
We perform high-resolution pseudo-spectral DNSs of \eq{eq:Burgers} on a
bi-periodic square domain with side $L = 2\pi$.
After the DNS reaches a non--equilibrium--statistically--stationary
state (NESS), we calculate the energy spectrum $E(k)$,
which shows a power-law regime that is consistent with 
$E(k)\sim k^{-5/3}$, over approximately one-and-half decades
[see \SMat{sm:dns}].
We also calculate the equal--time, longitudinal structure functions
$\Sp$ and their scaling exponents $\zetap$:
\begin{subequations}
  \begin{align}
    \Sp(r) &\equiv
    \avg{\lvert \dup(\rr)\rvert^p}\sim r^{\zetap}\/,
     \\
     \text{where}\quad \dup(\rr) &\equiv
                    \left[\uvec(\xx+\rr) - \uvec(\xx)\right]\cdot
      \left(\frac{\rr}{r}\right)\/.
  \end{align}
    \label{eq:Sp}
\end{subequations}
Here the symbol $\avg{\cdot}$ denotes averaging over the NESS.
Equation \eqref{eq:Burgers} exhibits
bi-scaling, given the type of forcing 
we use~\citep{Chekhlov,Chekhlov1995full}, i.e., the scaling exponents
 \begin{equation}
     \zetap =  \begin{cases}
     p/3 \quad\text{for}\quad p<3\/,\\
     1  \quad\text{for}\quad p \ge 3 \/.
     \end{cases}
     \label{eq:zetap}
\end{equation}
% for the forcing we use. 
 Our DNS results largely agree with this~\footnote{The numerical values of
     $\zetap$, measured
   from our DNS, deviate slightly from \eqref{eq:zetap} for
   $p \gtrsim 3$ (see \SMat{sm:dns}), possibly because of certain
   numerical artifacts identified in an earlier similar study in one
   dimension~\cite{BurgMitra}. Whether this deviation is a numerical
   artefact or not in $2D$ is not our main focus here.}.
 In the NESS, we seed the
 flow uniformly with tracers and track their subsequent motion.
 We present a pseudo-gray-scale plot of $\dive\uvec$ in  \subfig{fig:snap}{a},
 in which shocks are visible as dark filamentary structures.
 Tracers, shown in red, accumulate on these shocks.
 In \subfig{fig:snap}{b} and \subfig{fig:snap}{c} we give, respectively,
 illustrative surface plots of the $x-$ and $y-$ components of 
 $\uvec(\xvec,t)$ at the same instant of time as for \subfig{fig:snap}{a}.
 In these surface plots, the shocks appear as large negative jumps. 
 Henceforth, we non-dimensionalize all quantities by using $L$ and
 $\TL \equiv \LI/\urms$, the large-eddy-turnover time, with $\urms$ the
 root-mean-square velocity and $\LI$ the integral length
 scale [see \SMat{sm:dns} for details].

\begin{figure}[t]
    \centering
    \includegraphics[width=0.9\columnwidth]{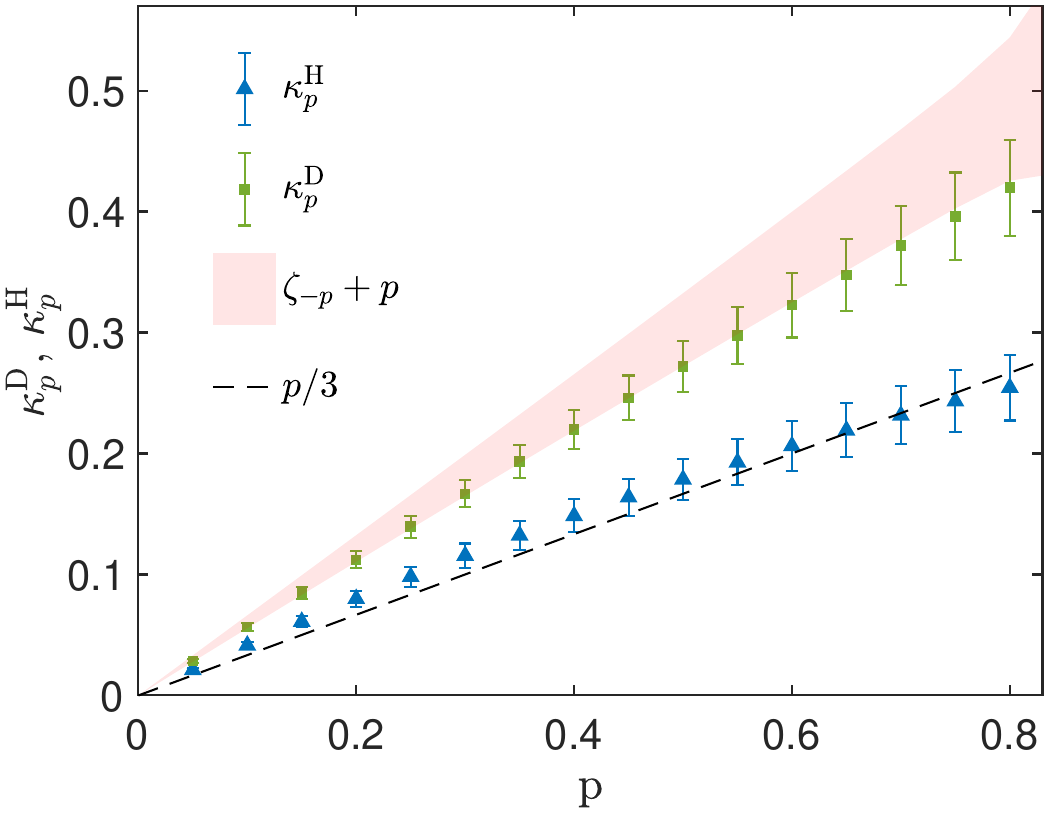}
    \caption{ \textbf{Dynamic scaling exponents}, of order $p$,
      as a function of $p$ for the
      doubling times $\kapDp$ (triangles)  and halving times
      $\kapHp$ (squares)  of Lagrangian intervals.
      Clearly, contrary to the naive expectation based on the conventional multifractal model,
      $\kapDp\ne\kapHp$.
      The region shaded in light pink is the multifractal
      prediction~\eq{eq:tauD}, where the equal-time scaling exponents
      $\zeta_{-p}$ and their error bars are calculated from our DNS.
      The dashed line is our theoretical prediction~\eq{eq:tauH}, which
      agrees very well with the numerical results. 
      }
    \label{fig:kapp}
\end{figure}

For a Lagrangian interval of initial length $R_0$, within the
inertial range, a naive application of the multifractal model of turbulence,
as done previously~\cite{Bofetta1999_Doubling,PairDisp_Bofetta2002, Bofetta2002_RichardIntermitt}, yields the following scaling exponents of the
order-$p$ moments of the doubling
times, $\tauD$:
\begin{subequations}
  \begin{align}
      \avg{\tauD^p(R_0)}&\sim R_0^{\kapDp}\,, \\
      \text{where}\quad\kapDp &= \zeta_{\rm -p}+p\,\label{eq:kapDp}\,.
      \end{align}
      \label{eq:tauD}
  \end{subequations}
  The second of these equations gives the bridge relation between the
  dynamic exponent, $\kapDp$, and the equal-time exponent, $\zeta_{\rm -p}$.
 We probe the validity of this bridge relation only for $p<1$,
 because the structure functions of negative order, $S_{-p}(r)$,
 exist only in this regime~\cite{Castaing1990Zerovel,Chen2005LowMom}. 
 We calculate $\kapDp$ and $\zeta_{-p}$, from our DNS
 and plot them versus $p$ in Fig.~\ref{fig:kapp}.
 Clearly \eq{eq:kapDp} holds within error bars.  
 By construction, the multifractal model does not distinguish between
 the scaling behaviors of doubling times and halving times, $\tauH$. 
  This may lead us to expect, for compressible turbulence,
  that the moments of $\tauH$ should have the same scaling as the moments
  of $\tauD$, i.e.,
  $\avg{\tauH^p(R_0)} \sim R_0^{\kapHp}$ with $\kapHp=\kapDp$.
  Our DNS, see \Fig{fig:kapp}, shows that this \textit{naive expectation
    is false}, because  $\kapDp\ne\kapHp$. 

%---------------------------------
%-------------------------

We must, therefore, generalize the multifractal model
in order to construct a theory that yields the $p$-dependence of $\kapHp$
shown in  \Fig{fig:kapp}.
We first outline the standard argument that is used to understand Richardson's
law~\citep{benzi2023lectures}.
A Lagrangian interval of length, $R(t)$, undergoes a Brownian
motion (we restrict ourselves to 2D) with a diffusivity, $K(R)$,
that depends on $R$ itself, i.e., the probability distribution
function (PDF) of $R$, namely, $W(R,t)$, satisfies
\begin{equation}
  \delt W = \frac{1}{R}\delR\left[R K(R)\delR W\right]\/.
    \label{eq:PairDiff}
\end{equation}
The $R$-dependence of $K(R)$ is deduced from the following dimensional
arguments: $K \sim (\dRV)^2\tcor $,
where $\dRV$ is the typical velocity difference
across the scale, $R$, and $\tcor$ is the typical correlation time.
If we now use the K41 forms,
$\dRV \sim R^{1/3}$ and
$\tcor \sim R/\dRV \sim R^{2/3}$, we get
$K \sim R^{4/3}$ 
which, when substituted in \eq{eq:PairDiff}, yields Richardson's
law~\footnote{An alternative argument is as follows.
By K41-type arguments, the only two quantities
than can appear in constructing $K$ are $\eps$ and $R$,
where $\eps$ is the energy dissipation rate per unit mass.
Hence $K \sim \eps^{1/3}R^{4/3}$.}.

To carry out a similar calculation for 2D Burgers
turbulence,
  we need an appropriate scaling form of $K(R)$.
  The first, straightforward, generalization is to replace
  the K41 result, $\dRV \sim R^{1/3}$, with   
  $\dRV \sim R^{h}$, where $h$ is the scaling exponent of the velocity field. 
  The second key idea is to recognize that, when we consider a Lagrangian
interval of length $R$ here,
%in 2D Burgers equation, 
we must distinguish
between the following three possibilities for this interval:
\begin{enumerate}
\item Case (1) [interval along a shock]: 
  $\dRV \sim R^{h}$, but the  typical correlation time is determined by sweeping as both ends of the interval
are trapped in the same shock, so $\tcor \sim R$, whence
$K \sim R^{2h+1} $.
\item Case (2) [interval straddles a shock]: $\dRV$ is a constant,
  independent of $R$, i.e., $h=0$, and $\tcor \sim R$, consequently $K \sim R$. 
\item Case (3) [interval is away from shocks]: The interval can decrease in
  only one of the following two ways:
\begin{enumerate}[(A)]
\item Case (3A): Both ends of the interval get trapped,
  at a later time, in the same shock, so the arguments used in case (1) apply.
  Therefore, at late times, $K \sim R^{2h+1}$. 
\item Case (3B): The two ends of the interval get trapped in two
different shocks, which then approach each other.
Hence, $\dRV$ is then the velocity difference between the two shocks, which does not depend on $R$. 
Since $\tcor \sim R$, this yields $K\sim R$ at late times. 
\end{enumerate}
\end{enumerate}
In all of these cases, the  calculation of the PDF
of halving times is a first-passage problem for \eq{eq:PairDiff}, with two
boundaries, one at $R\to \infty$ and the other at $R=\Rnot/2$,
where $\Rnot\equiv R(t=0)$.
Given the forms of $K(R)$ in Cases (1)-(3), 
it is sufficient to calculate the halving time PDFs, $\Pone$ and $\Ptwo$, for
$K \sim R^{2h+1}$ and $K \sim R$, respectively.
We obtain~\citep{Redner2001guide}
\begin{subequations}
  \begin{align}
    \Pone(\tauH,\Rnot) &\sim\frac{1}{\Rnot^{1-2h}}
    \exp\left[-\Aone\frac{\tauH}{\Rnot^{1-2h}} \right] \quad\text{and} \\
    \Ptwo(\tauH,\Rnot) &\sim\frac{1}{\Rnot}
                 \exp\left[-\Atwo\frac{\tauH}{\Rnot} \right]\/, 
  \end{align}
  \label{eq:P12}
\end{subequations}
where $\Aone$ and $\Atwo$ are numerical constants~[see  
  \SMat{sm:theory} for details]. By collecting all the cases together, we find
the overall PDF of $\tauH$, for large $\tauH$:
\begin{equation}
  \mP(\tauH,\Rnot) \sim \wone\Pone + \wtwo\Ptwo + \wthA\Pone + \wthB\Ptwo\,,
  \label{eq:PP}
\end{equation}
where $\wone$, $\wtwo$, $\wthA$, and $\wthB$ are the relative weights
for the Cases discussed above.
As the shocks are one-dimensional structures, $\wone$ is negligible,
but the other weights are not; and in the limit $\Rnot \to 0$,
$\wthA\Pone$ is the dominant contribution.
Thereby, we obtain the following result for the moments of the PDF of $\tauH$:
\begin{equation}
  \avg{\tauH^p} \sim \Rnot^{p(1-2h)} \sim \Rnot^{p/3}\/.
  \label{eq:tauH}
  \end{equation}
In the final step we substitute $h=1/3$~\citep{Chekhlov,Chekhlov1995full}. 
Equation~\ref{eq:tauH} agrees well with our numerical 
results~[see the dashed line in \Fig{fig:kapp}]. 

Several important comments are now in order:
\begin{enumerate}[(a)]
\item We emphasize that, although the halving-time dynamic exponent
  $\kapHp$ depends linearly on the order $p$,
  the statistics of halving times is \textit{not Gaussian}
  -- the tail of the PDF of $\tauH$ is exponential [see \eq{eq:P12} and
  \eq{eq:PP}].
  Furthermore, \eq{eq:tauH}  \textit{does not} imply simple dynamic scaling
  because we have shown explicitly that differently defined time-scales
  have different scaling exponents.
\item Although we discuss halving and doubling
  times, explicitly, our results apply to exit times in general. 
\item Even though our DNS study is limited to the randomly-forced
  2D Burgers equation~\ref{eq:Burgers}, our theoretical arguments can be
  generalized to other, richer varieties of shock-dominated turbulence 
  in a straightforward manner, as long as the
  equal-time scaling exponents, $\zetap$, are known.
\end{enumerate}

%----------------------------------------
\begin{figure}[t]
    \centering
    \includegraphics[width=0.9\columnwidth]{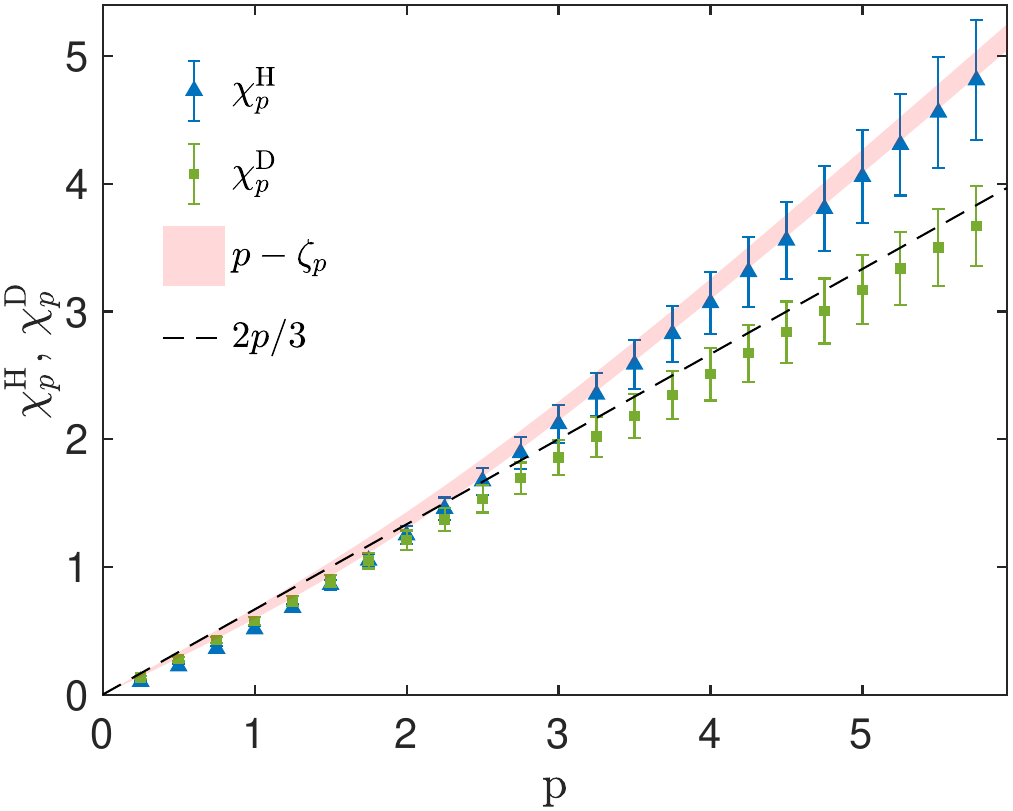}
    \caption{\textbf{Dynamic scaling exponents}, of order $p$
      as a function of $p$ for the
      doubling frequencies $\chiDp$ (squares)  and halving frequencies
      $\chiHp$ (triangles)  of Lagrangian intervals.
      Clearly, contrary to the naive expectation based on the conventional multifractal model,
      $\chiDp\ne\chiHp$ for all $p$. 
      The region shaded in light pink is the predicted bridge relation
      $\chiHp=p-\zetap$, where the equal-time
      scaling exponents $\zetap$ and their error bars are calculated from
      our DNS.  
      The dashed line is our prediction for $\chiDp$.      
      }
    \label{fig:chip}
\end{figure}
%--------------------------------------
 A critical problem with the multifractal description of the
 scaling of exit times is that the bridge relation~\ref{eq:tauD}, 
 requires the calculation of negative moments of $\dRV$.
 One way to avoid the calculation of such moments is to consider
 the statistics of \textit{inverse} exit times. 
 In particular, we define the halving and doubling frequencies
 to be $\omH \equiv 1/\tauH$ and $\omD \equiv 1/\tauD$,
 respectively. 
 Thereby we define two new sets of dynamic scaling exponents,
 \begin{equation}
     \avg{\omH^p(R_0)}\sim R_0^{-\chiHp} \quad\text{and}\quad
     \avg{\omD^p(R_0)}\sim R_0^{-\chiDp}\,.
     \label{eq:chip}
 \end{equation}
 As we have discussed already, the  multifractal model for incompressible turbulence
 does not distinguish between them~\cite{Bofetta1999_Doubling,
   PairDisp_Bofetta2002, Bofetta2002_RichardIntermitt} so the naive expectation is
 $\chiHp=\chiDp=p-\zetap$. 
 On the contrary, our DNS shows (within the error bars of \Fig{fig:chip}], 
 \begin{equation}
   \chiHp = p - \zetap\quad\text{and}\quad \chiDp = 2p/3\/;
 \end{equation}
 clearly, $\chiHp\ne\chiDp$. 
 Our extension of the multifractal model predicts the correct
 exponents in the following manner. 
 The scaling of the moments of $\omD$ and $\omH$, for  $p>1$,
 is determined  primarily by the small-$\tauD$ and small-$\tauH$ behaviors
 of their respective PDFs.
 The  short-time growth of $R_0$ can occur only if the
 interval is away from the shocks [Case (3), but at short times].
 Therefore, for all such intervals, $K \sim R^{4/3}$ and consequently
 $\chiDp = 2p/3$ for all $p$ [dashed line in \Fig{fig:chip}],
 in agreement with the results of our DNS. 
 As for $\chiHp$, at short times, even for incompressible turbulence,
 intervals can decrease, hence the the prediction of the multifractal
 model holds.
  
In an earlier paper~\citep{De_DynScal} we have investigated dynamic
multiscaling in the 1D stochastically forced Burgers
equation by using a slightly different formulation
than we have used here.
Instead of halving times, we have considered interval-collapse times, i.e.,
the time it takes for a Lagrangian interval to shrink to a point.
We have shown that, in 1D, the interval-collapse-time PDF has power-law tails.
In contrast, we now find that, in 2D, the PDF of halving times has exponential
tails.
In both 1D and 2D cases, the shocks play a key role in determining the
multiscaling properties.

 In summary, we have shown how to generalize the multifractal model for
 incompressible-fluid turbulence to the stochastically forced
 2D Burgers equation, which is rich enough to display the complexities
 of shock-dominated turbulence.
 Our DNS demonstrates clearly that the statistics of halving times
 deviate starkly from those of doubling times.
 By generalizing the standard argument that is used to derive
 Richardson's law from K41 theory, we have developed a theory that leads to
 a natural way of understanding the statistics of halving and doubling times.
 The key idea in this theory is that, when we consider a Lagrangian interval of
 length $R$, we must distinguish between
 Cases (1)-(3).
 Our theoretical arguments can be generalized to
 shock-dominated compressible turbulence, if the equal-time
 exponents $\zetap$ are known.

We expect shock--dominated turbulence in astrophysical systems
at both high Mach and Reynolds numbers, e.g., in
the interstellar media and molecular clouds, where the turbulence is driven
by supernovae explosions.
The dynamics of Lagrangian intervals in such flows provides useful insights
into transport and mixing in these systems, which  influence chemical kinetics
and the rates of formation of stars and
planetesimals~\cite{Mordecai2004Review}. 
We expect our generalization of Richardson's law to apply to such
systems.
For compressible turbulence, where the  irrotational and
solenoidal components of the flow have similar energies, our
theory may require further generalization.
Our work brings out the importance of calculating the statistics of both
halving and doubling times of Lagrangian intervals from DNSs of compressible
turbulent flows, for trans-sonic, super-sonic, and hyper-sonic cases.

\acknowledgements
DM acknowledges the financial support of the Swedish Research Council
through grants  638-2013-9243 and 2016-05225.
NORDITA is partially supported by NordForsk. SD thanks the PMRF (India) for
support. RP and SD thank the Science and Engineering Research Board (SERB) and the
National Supercomputing Mission (NSM), India for support, and the
Supercomputer Education and Research Centre (IISc) for computational resources.

% %-----------------------------------

%\bibliography{burg,ref}

\newpage
\onecolumngrid
\appendix
\section{Direct Numerical Simulations}
\label{sm:dns}
We perform pseudospectral direct numerical simulations (DNSs) of the randomly
forced 2D Burgers equation
\begin{eqnarray}
  \delt\uvec + \uvec\cdot\nabla\uvec &=&
  \nu\nabla^2\uvec + \fvec(\xvec,t)\,;
  \quad\nabla\times\uvec=0\,; \nonumber \\
        \avg{\fhat(\kvec,t)\cdot\fhat(\kvec',t')}
&\sim& |\kvec|^{-2}\delta(\kvec+\kvec')\delta(t-t')\,;
    \label{sm:Burgers}
\end{eqnarray}
on a square, doubly-periodic domain of side length $L=2\pi$ and with $N^2$ collocation points. We use the $2/3$-dealiasing~\citep{rog81,Can88} scheme to remove aliasing errors and the second-order exponential time-differencing Runge-Kutta (ETDRK2) scheme~\citep{cox+mat02} for the time evolution of the velocity field. The stochastic force is incorporated  by using the Euler-Maruyama scheme~\citep{hig01}. The parameters of our DNS runs are given in Table~\ref{tab:parameters}.

%--------------------------------------------%
\begin{figure*}[h]
    \centering
    \includegraphics[width=0.9\columnwidth]{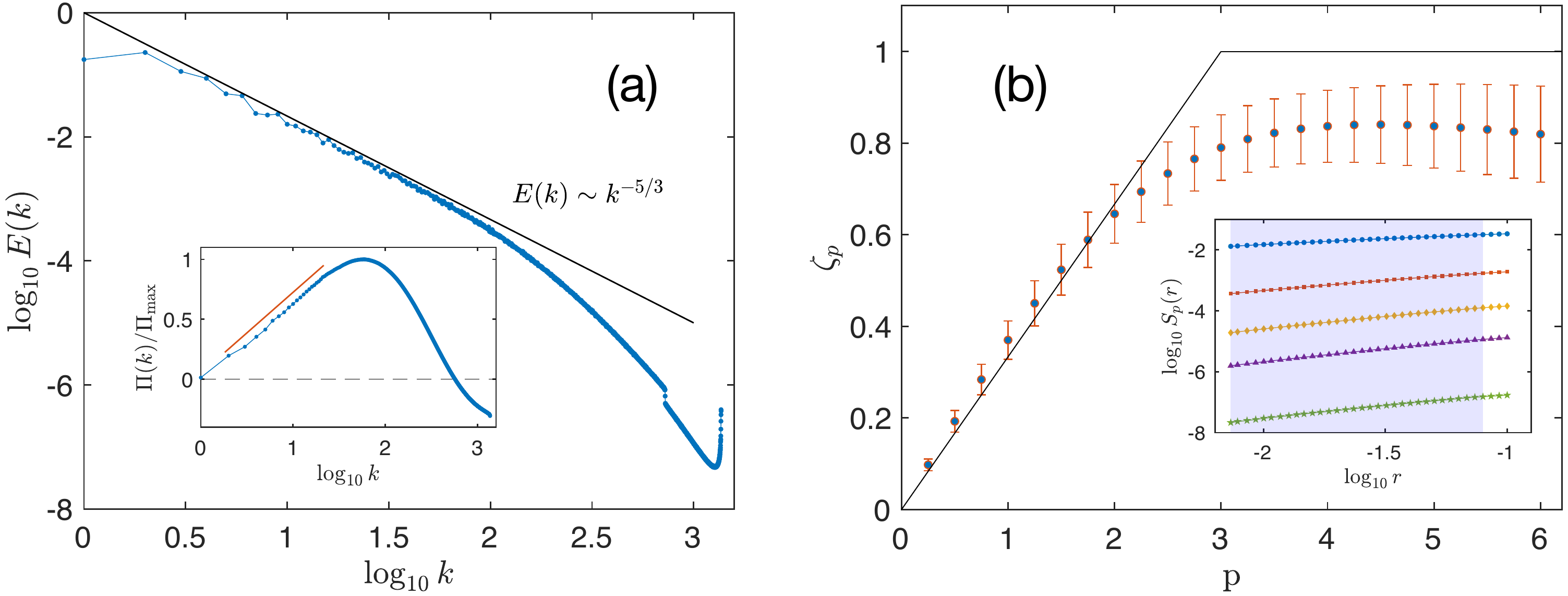}
    \caption{\small (a) Log-log plot versus the wave number $k$ of the energy spectrum $E(k)$ (averaged over shells of radius $k$
    in Fourier space and over time in the nonequilibrium statistically steady turbulent state);
      the inertial range extends over nearly a decade and a half in $k$,
      with $E(k)\sim k^{-5/3}$ (black line).
      \textit{Inset:} Net energy flux $\Pi(k)$ through the wave-number scale $k$,
      displaying a direct cascade of energy; the red straight line indicates
      $\Pi(k)\sim A \log k$, where $A$ is a constant,
      as suggested on theoretical grounds for the randomly-forced 1D Burgers and
      3D randomly forced Navier-Stokes turbulence~\cite{BurgMitra,Sain1998}.
      (b) Equal-time structure function exponents, $\zetap$, versus $p$;
      the black lines represent bifractal scaling.
      \textit{Inset:} Log-log plots of the Eulerian equal-time structure
      functions, $\Sp(r)$, versus $r$ for $p=1$ (circles), $p=2$ (squares),
      $p=3$ (triangles), $p=4$ (triangles) and $p=6$ (stars); the
      shaded region denotes the regime of best power-law fit within the
      inertial regime;
      $\zetap$ are calculated via local slope analyses of these curves.}
    \label{fig:Fig1app}
\end{figure*}
%------------------------------------%
\begin{table*}[h]
\centering{}
\begin{tabular}{c c c c c c c c c}
\hline
$N$ & $\nu$ & $\delta t$ & $L_I$ & $\urms$ & Re & $\eta$ & $T_L$ & $k_{\rm max}\eta$\\
\hline
\hline
$4096$ & $5\times10^{-5}$ & $5\times10^{-4}$ & $0.25$ & $0.05$ & $1.61\times10^3$ & $5.6\times10^{-3}$ & $30.0$ & $7.7$\\
\hline
\end{tabular}
\caption{\label{tab:parameters}{Parameters for our direct numerical
    simulations (DNS): $N$ is the number of collocation points,
    $\nu$ the kinematic viscosity, $\delta t$ the time step,
    $\LI\equiv\frac{3\pi}{4}\frac{\int\frac{E(k)}{k}dk}{\int E(k)dk}$
    the integral length scale (normalized by $2\pi$ here),
    $\urms$ the root-mean-square velocity, ${\rm Re}=\urms \LI/\nu$
    the integral-scale Reynolds number,
    $\eta\equiv\left(\frac{\nu^3}{\epsilon}\right)^{1/4}$ the dissipation
    length scale, where $\epsilon$ is the mean energy dissipation rate,
    $\TL\equiv\LI/\urms$ the large-eddy-turnover time, and $\kmax$ the
    de-aliasing cutoff.}}
\end{table*}
%------------------------

Figure~\ref{fig:Fig1app}(a) displays the ensemble-averaged energy spectrum,
\begin{equation}
    E(k) \equiv {1\over2}\sum_{|\kvec|=k}|\hat{\uvec}(\kvec)|^2\,,
    \label{sm:E}
\end{equation}
in the non-equilibrium statistically steady state (NESS).
In the inertial range, the power-law behavior of the energy specturm is consistent
with $E(k)\sim k^{-5/3}$ over more than a decade and
half~\cite{BurgMitra,De_DynScal} in wavenumber $k$.
The scale-to-scale energy flux, $\Pi(k)$, through the scale $k$ is not
constant in the inertial range but has logarithmic corrections,
i.e., $\Pi(k)\sim\log k$, which is agreement with earlier results in randomly
forced turbulence wherein all modes up to a cut-off, $\kc\gg k$
(with $k$ in the inertial range), are forced randomly [see Refs.~\cite{Sain1998,BurgMitra}
for details]; we choose $\kc=\sqrt{2}N/8$.
We calculate the longitudinal equal-time structure functions,
\begin{subequations}
  \begin{align}
  \Sp(r)&=\avg{\lvert\delta\uvec(\rr)\cdot \hat{r}\rvert^p}
  \sim r^{\zetap}\/,\quad\text{where}\\
  \delta\uvec(\rr)&=\uvec(\xvec+\rr)-\uvec(\xvec)\/.
 \end{align}
\label{sm:Spr}
\end{subequations}
Here $\hat{r}$ is the unit vector in the direction of $\rr$.
We plot $\Sp(r)$ in the inset of \subfig{fig:Fig1app}{b} on a log-log graph;
in the inertial range, $\eta/L\ll r\ll1$, we find clean power-law scaling for $\Sp(r)$; by
carrying out local slope analyses, we evaluate the exponents $\zetap$ that we show in
\subfig{fig:Fig1app}{b}.
These exponents are close to the bifractal scaling predicted for stochastically forced 1D Burgers turbulence~\cite{Chekhlov1995full,BurgMitra};
whether their deviation from the bifractal prediction for $p\geq3$
signifies bonafide multiscaling (or is simply a numerical artifact) requires
in-depth theoretical and numerical analysis, both of which lie beyond
the scope of this article.

The equation of motion of a  Lagrangian particle or tracer is 
\begin{equation}
    \frac{d}{dt}\Xvec(t)=\uvec(\Xvec,t)\,,
    \label{sm:tracer}
\end{equation}
where $\Xvec(t)$ is the position of the tracer at time $t$.
After the flow has reached the NESS, we seed the flow uniformly with
$\Np=2048^2$ tracers on a square grid; subsequently, we track their motion by
solving \eqref{sm:tracer}, where we calculate $\uvec(\Xvec,t)$ via bilinear
interpolation of the Eulerian velocities~\citep{yeu89} to off-grid points. 

\section{Theory for halving time of Lagrangian intervals}
\label{sm:theory}
We now outline the steps that are necessary to model the pair diffusivity $K(R)$
of a Lagrangian interval of size $R$. Once we have $K(R)$ we derive, from the corresponding Fokker-Planck
equations, the asymptotic forms of the PDFs of the halving times of the
Lagrangian intervals with initial size $R_0$, that lies
in the inertial range [see Eq. (6) in the main text]. 

The velocity difference $\dRV$, across an interval of length $R$, is either
\begin{enumerate}
\item  a constant independent of $R$, when the interval straddles a shock,
\item or  $\dRV\sim R^h$, where 
$h$ is the local H\"older exponent of
the velocity field; for the type of forcing we use, 
$h=1/3$~\citep[see e.g.,][]{Chekhlov,Chekhlov1995full}.
\end{enumerate}
For a Lagrangian interval one of the following
three cases must hold:
\begin{enumerate}
\item  Case (1): the interval lies along a shock.
\item  Case (2): the interval straddles a shock. 
\item Case (3): the interval lies away from any shock.
\end{enumerate}
There is the additional possibility that the two ends of
the interval are in two different shocks. We show later that it is not necessary to 
consider this case separately.
 
%----------------------------------
\begin{figure}[h]
    \centering
    \includegraphics[width=0.4\columnwidth]{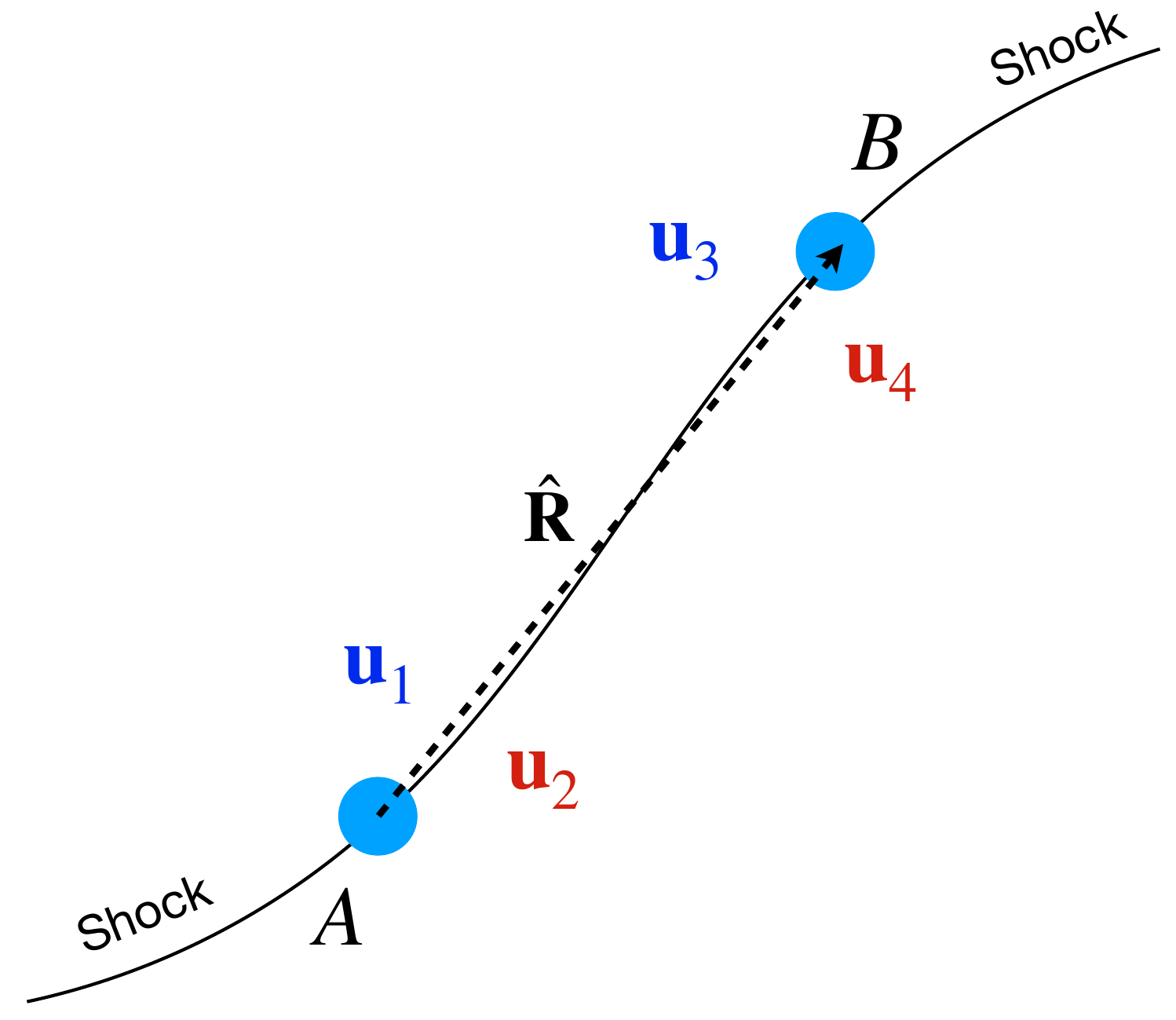}
    \caption{\small Schematic diagram of a typical case of a Lagrangian interval of length $R$ 
lying along a shock (solid line). $A$ and $B$ marks the two ends of the  interval.} 
    \label{fig:PairOnShock}
\end{figure}
%----------------------
Case (1) [see \Fig{fig:PairOnShock}]: 
Then the velocities at $A$ and $B$ are $(\uvec_{\rm 1}+\uvec_{\rm 2})/2$ and
$(\uvec_{\rm 3}+\uvec_{\rm 4})/2$, respectively. 
Hence, 
\begin{equation}
\dRV = (\uek-\utin)\cdot\hat{R}+(\udo-\uchar)\cdot\hat{R}
\label{eq:dR1}
\end{equation} 
As both the velocity differences on the right-hand-side  of \eq{eq:dR1}
scale as $R^h$,  $\dRV\sim R^h$. 
Furthermore, because both the ends of the interval are on the same  shock, 
the typical correlation time $\tcor\sim R$ because of the sweeping effect. 
Therefore, $K\sim R^{1+2h}$.

Case (2): $\dRV$ is a constant independent of $R$ and $\tcor \sim R$.
Consequently, $K \sim R$. 

Case (3): the interval can decrease in the following two ways.
\begin{enumerate}
\item Case (3A): both ends of the interval get trapped, at a later time, in the same shock, so
the arguments used in Case (1) apply; but, after the interval is trapped on the shock, $K \sim R^{2h+1}$ at
late times. 
\item Case (3B): at late times the two ends of the interval get trapped in two
different shocks which then approach each other. The velocity difference $\dRV$ is the velocity difference
between two shocks; this does not depend on the distance between them,
i.e., $\dRV$ is independent of $R$; furthermore, $\tcor \sim R$,
so $K \sim R$.
\end{enumerate}
  
In all of these cases, the calculation of the PDF of halving times is a first-passage time problem 
for the following 2D Fokker-Planck equation for the the PDF $W(R,t)$ of $R$:
\begin{equation}
    \partial_t W=\frac{1}{R}\partial_R[RK(R)\partial_R]W\,,
    \label{sm:Eq1}
\end{equation}
with absorbing boundaries at $R\to\infty$ and $R=R_0/2\equiv R(t=0)/2$. 
It is sufficient to do such a calculation for the two cases with
$K \sim R^{2h+1}$ and $K \sim R$.
We make the change of variable $\xi=1/R$, 
which maps the radial domain to the interval $[0,2/R_0]$. 
The Fokker-Planck equation in $\xi$ then reads,
\begin{equation}
  \frac{\partial}{\partial t}W(\xi,t) =
  K_0\xi^3\frac{\partial}{\partial\xi}
  \left[\xi K(\xi)\frac{\partial}{\partial\xi}W(\xi,t)\right]\,.
    \label{sm:Eq1a}
\end{equation}

We first consider the case $K\sim R^{1+2h}$, i.e., $K=K_0\xi^{-1-2h}$, where $K_0$
is an $R-$independent constant. We define $y=\xi^{(2h-1)/2}$ and substitute it in \eqref{sm:Eq1a}
to obtain
\begin{equation}
  \frac{\partial}{\partial t}W(y,\tau)
  =K_0\left(\frac{1-2h}{2}\right)^2
  \left[\frac{\partial^2}{\partial y^2}+\left(\frac{3+2h}{1-2h}\right)
    \frac{1}{y}\frac{\partial}{\partial y}\right]W(y,t)\,.
    \label{sm:Eq2}
\end{equation}
We solve \eqref{sm:Eq2} with the boundary condition $W(y=Y,t)=0$, where
$Y=(2/R_0)^{(2h-1)/2}$, by separation of variables. 
We obtain
\begin{equation}
  W(y,t)=\sum_{n}\frac{A_n}{y^q}J_q\left(\frac{c_ny}{Y}\right)
  \exp\left[-c_n^2K_0\left(\frac{1-2h}{2}\right)^2
    \frac{t}{Y^2}\right]\,,
    \label{sm:Eq3}
\end{equation}
where $J_q$ is the order-$q$ Bessel function of the first kind,
where $q=(1+2h)\big/(1-2h)$, $c_n$ is the $n-$th zero of
$J_q$, and $A_n$ is the normalization constant.

The initial distribution is
$W(\xi,t=0)=W_0(\xi)=\delta(\xi-\xi_{\ast})/2\pi\xi$, where $\xi_\ast=1/R_0$.
In the variable $y$ we have
\begin{equation}
  W_0(y)=\frac{1}{2\pi y^{2/(2h-1)}}
  \left(\frac{1-2h}{2}\right)
  \left(\sigma Y\right)^{(3-2h)/(1-2h)}\delta (y-\sigma Y)\,,
\end{equation}
where $\sigma={1/2}^{(2h-1)/2}$. We set $t=0$, multiply both sides of \eqref{sm:Eq3} by 
$y^{q+1}J_q(c_my/Y)$, integrate from $y=0$ to $y=Y\sigma$, and by
using the orthogonality properties of Bessel functions, we get
\begin{equation}
    A_m=\frac{1-2h}{4\pi}\left(\sigma Y\right)^{(5+2h)/(1-2h)}\frac{J_q(c_m/\sigma)}{J_{q+1}(c_m)}\,,
\end{equation}
for any positive integer $m$. Thus,
\begin{equation}
    W(y,t)=\frac{1-2h}{4\pi}\left(\sigma Y\right)^{(5+2h)/(1-2h)}\frac{1}{y^q}\sum_{n}\frac{J_q(c_n/\sigma)}{J_{q+1}(c_n)}J_q\left(\frac{c_ny}{Y}\right)\exp\left[-c_n^2K_0\left(\frac{1-2h}{2}\right)^2\frac{t}{Y^2}\right]\,.
    \label{sm:Eq4}
\end{equation}
The survival probability is given by
\begin{equation}
    \mathcal{S}(\Rnot,t)=\int_{\xi<2/R_0}W(\xi,t)\xi d\xi d\theta\,.
\end{equation}
In the new variable $y$, the area element
$\xi d\xi d\theta=\frac{2}{1-2h}y^{(2h-5)/(1-2h)}dy d\theta$, so
\begin{equation}
  \mathcal{S}(Y,t)=
  \left(\sigma Y\right)^{(5+2h)/(1-2h)}
  \sum_{n}\frac{J_q(c_n/\sigma)}{J_{q+1}(c_n)}
  \exp\left[-c_n^2K_0\left(\frac{1-2h}{2}\right)^2
    \frac{t}{Y^2}\right]\int_0^{Y} y^{1+q}J_q
  \left(\frac{c_ny}{Y}\right)dy\,.
\end{equation}
For large $t$, the $n=1$ term dominates  over the others.
Hence, by retaining only its contribution and substituting $z=y/Y$ in the
integrand, we get
\begin{equation}
  \mathcal{S}(Y,t)\simeq G(\sigma)
  \exp\left[-c_1^2K_0\left(\frac{1-2h}{2}\right)^2
    \frac{t}{Y^2}\right]\,,
  \quad\text{where}
  \quad G(\sigma)=\sigma^{(5+2h)/(1-2h)}
  \frac{J_q(c_1/\sigma)}{J_{q+1}(c_1)}\int_0^1 z^{q+1}J_q(c_1z)dz\,.
    \label{sm:Eq5}
\end{equation}
By replacing $Y$ by $(2/R_0)^{(2h-1)/2}$, we can evaluate the halving-time distribution $\Pone(\tauH,\Rnot)$ as
follows:
\begin{equation}
    \Pone(\tauH,\Rnot)=-\frac{d}{d\tau}\mathcal{S}(\Rnot,\tauH)\Bigg|_{t=\tauH}\,,
\end{equation}
whence we obtain the following asymptotic form:
\begin{equation}
    \Pone(\tauH,\Rnot) \sim\frac{1}{\Rnot^{1-2h}}
    \exp\left[-\Aone\frac{\tauH}{\Rnot^{1-2h}} \right]\,,
    \label{sm:Eq6}
\end{equation}
where $\Aone$ is a numerical constant.

Similarly, for the intervals with $K\sim R$, we obtain the asymptotic form of their halving-time distribution:
\begin{equation}
    \Ptwo(\tauH,\Rnot) \sim\frac{1}{\Rnot}
    \exp\left[-\Atwo\frac{\tauH}{\Rnot} \right]\,.
    \label{sm:Eq7}
\end{equation}

For Cases (3A) and (3B), let $\tau_{\rm s}$ be the time at which both ends of the interval get trapped on 
the same or different shocks, respectively. Then, for $\tauH\gg\tau_{\rm s}$, the halving-time distributions 
for Cases (3A) and (3B) reduce to $\Pone$ and $\Ptwo$, respectively. 
Therefore, the overall halving-time PDF $\mP(\tauH,\Rnot)$, for an interval of initial size $R_0$, 
is a weighted sum of all these cases [see Eq. (7) in the main text]. 
Thus, by retaining only the leading-order contribution in the limit $\Rnot\ll1$, we get
\begin{equation}
    \mP(\tauH,\Rnot)\sim \Pone \sim\frac{1}{\Rnot^{1-2h}}
    \exp\left[-\Aone\frac{\tauH}{\Rnot^{1-2h}} \right]\,,
    \label{sm:Eq8}
\end{equation}
which yields Eq. (8) in the main text.

\begin{figure}[h]
    \centering
    \includegraphics[width=0.6\columnwidth]{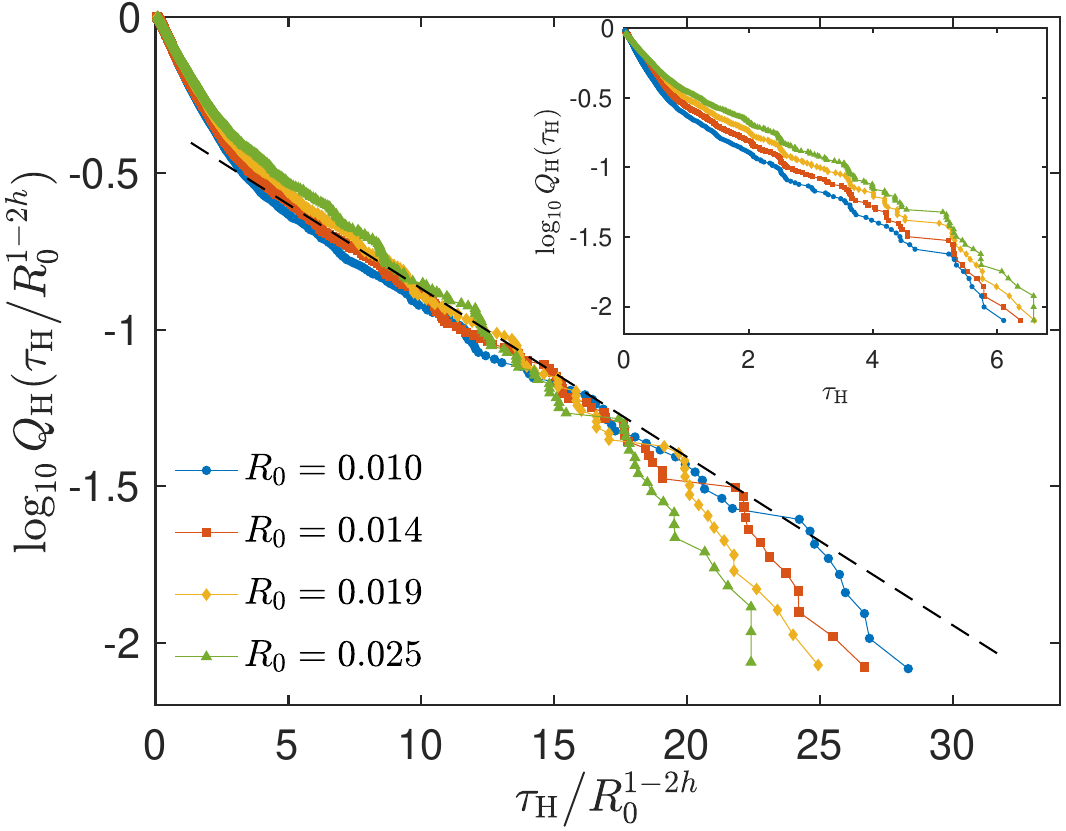}
    \caption{\textbf{Cumulative PDF of halving times}
The inset shows the semi-log plots of the cumulative PDFs (CPDFs) $Q(\tauH)$
of halving times. The main figure shows the collapse of the PDFs,  
after $\tauH$ is scaled by $R_0^{1-2h}$ (see text for our theoretical result). 
The dashed straight line is drawn for reference. The main figure and the
inset use the same colors and symbols. 
}
\label{fig:thalf_cpdf}
\end{figure}

To validate \eqref{sm:Eq8}, we evaluate the halving-time cumulative PDFs $Q(\tauH)$ from our DNS, 
for different initial interval sizes $R_0$, by using the rank-order method, which ensures that the CPDfs are free of binning errors. According to \eqref{sm:Eq8} this CPDF scales as
\begin{equation}
    Q(\tauH,R_0)\sim \exp\left[-\Aone\frac{\tauH}{R_0^{1-2h}} \right]\sim \exp\left[-\Aone\frac{\tauH}{R_0^{1-2h}} \right]\,.
    \label{sm:Eq9}
\end{equation}
In \Fig{fig:thalf_cpdf}, we plot $Q(\tauH)$, for different values of $R_0$ in
the inertial range, on a semilog graph. All the plots are consistent with exponential tails, 
which collapse onto a single line (shown dashed in the figure) on rescaling $\tauH$ by $R_0^{1-2h}$; this
confirms our theoretical result. 

%--------------------------------
%------------------------------------

\begin{figure}[h]
    \centering
    \includegraphics[width=0.6\columnwidth]{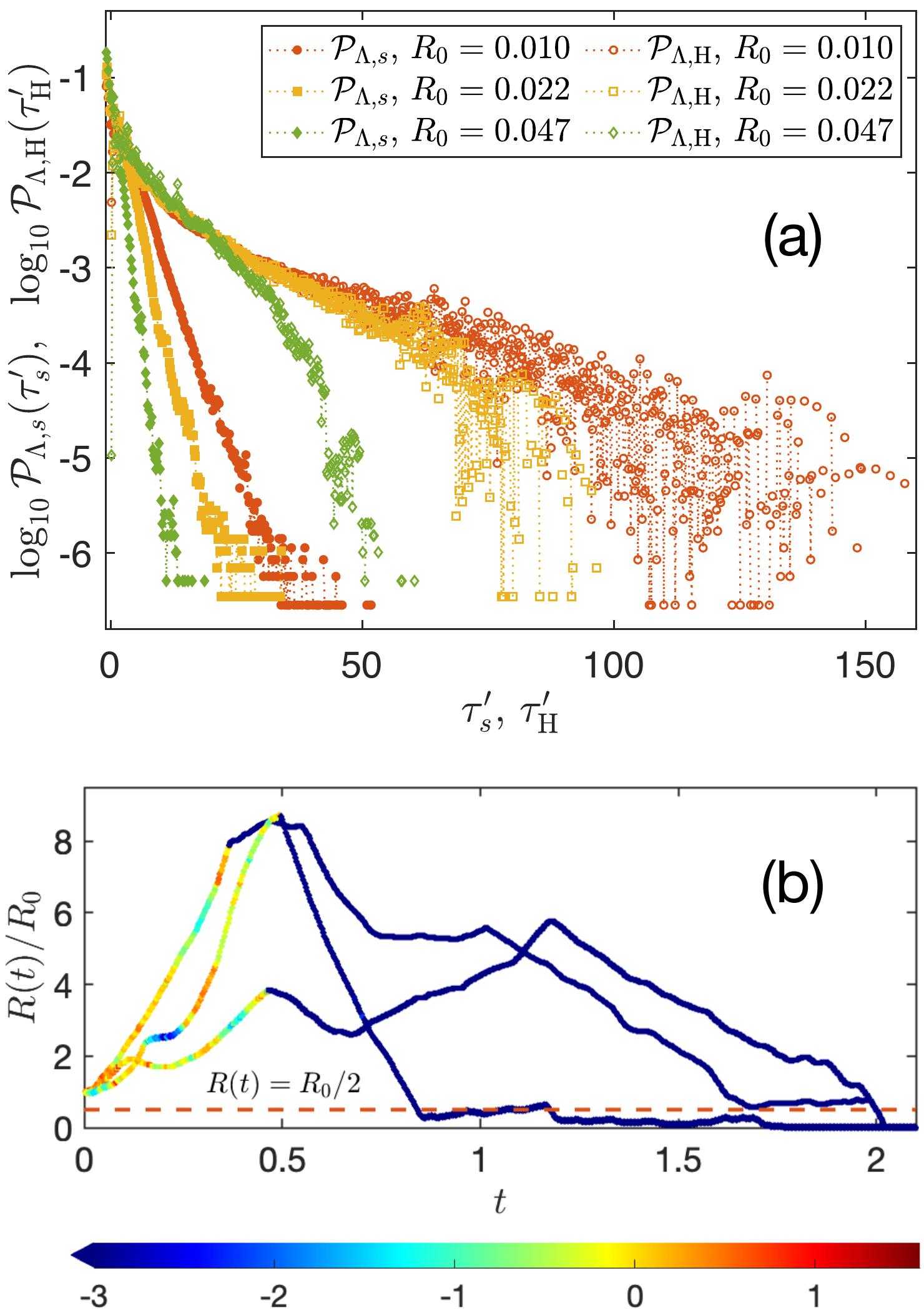}
    \caption{\small (a) Semilog plots of $\PLams(\tau_s')$ and
      $\PLamH(\tauH')$ for different values of $R_0$. This indicates that
      $\tauH$ is generally much greater than $\tau_s$ for a typical
      interval belonging to Cases (3A) or (3B), for any $R_0$.
      (b) Representative time series of interval sizes, $R(t)$, for $R_0=0.01$; colors
      denote the mean of the values of $\nabla \cdot \bm u$ at the two ends of the interval $R(t)$.}
    \label{fig:PairExit_DivTs}
\end{figure}

Furthermore, to verify the assumption that $\tauH\gg\tau_{\rm s}$ for
intervals belonging to Cases (3A) and (3B), we evaluate the following numerically:
\begin{enumerate}
\item The halving-time PDFs $\PLamH$, of intervals whose lengths initially
exceed a fixed threshold, $\Lambda=1.5R_0$, before contracting.
\item The PDFs $\PLams$, of $\tau_{\rm s}$ for these intervals. 
\end{enumerate}
The intervals in Cases (3A) and (3B) lie in the smooth regions of the flow at $t=0$,
so, at short times, they grow; and, as they do so, their ends fall on shocks which 
eventually lead to their contraction. Therefore, $\PLamH$ and $\PLams$ are representative PDFs of 
$\tauH$ and $\tau_{\rm s}$, respectively, for these intervals. 
We define $\taua$ to be the time at which such an interval size exceeds $\Lambda$ and
plot $\PLamH$ and $\PLams$ as functions of $\tauH'=(\tauH-\taua)/\taua$
and $\tau_{\rm s'}=(\tau_{\rm s}-\taua)/\taua$, respectively,  for different values
of $R$, in \subfig{fig:PairExit_DivTs}{a}. We observe that the tails of $\PLamH$ are much broader 
than those of $\PLams$ for every value of $R_0$. 
This provides compelling evidence in favor of our assumption that, generally 
$\tauH\gg\tau_{\rm s}$, for intervals belonging to Cases (3A) and (3B). 
For the purpose of visualization, we plot some representative time
series of sizes, $R(t)$, of these intervals, coloured by the respective means of the 
instantaneous velocity
divergences at their ends in \subfig{fig:PairExit_DivTs}{b}.
%-------------------------------
%-------------------------------------

\begin{figure}[h]
    \centering
    \includegraphics[width=0.95\columnwidth]{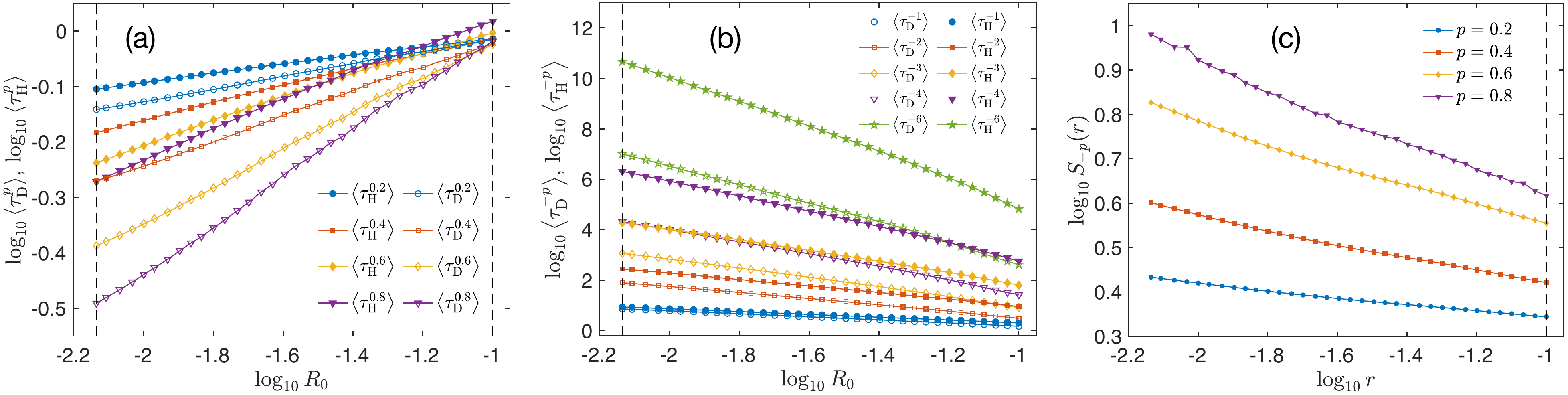}
    \caption{\small Log-log plots of some of the (a) positive fractional moments (whose scaling exponents are shown in Fig. 2 in the main text) and (b) negative moments (whose scaling exponents are shown in Fig. 3 in the main text) of $\tauH$ and $\tauD$ as functions of the initial interval size $R_0$. (c) Negative-order equal-time structure functions $S_{-p}(r)$ for different values of $p$; the scaling exponents $\zeta_{-p}$ are also shown in Fig. 2 in the main text. In all these figures, the vertical lines indicate the boundaries of the regimes of local-slope analyses, which yield the various scaling exponents and their error bars. For all cases, the scaling regimes extend over more than a decade in the independent variable.}
    \label{fig:MomData}
\end{figure}

In Fig.~\ref{fig:MomData} we present plots of the dependence of the moments of halving and doubling times on the initial interval size; we also give plots of negative-order equal-time structure functions.

\end{document}